\documentstyle[12pt,epsf]{article}
\def\thebib#1{
 \list
 {\arabic{enumi}}{\settowidth\labelwidth{[#1]}\leftmargin\labelwidth
 \advance
 \leftmargin\labelsep
 \setlength{\parsep}{0mm}%
 \setlength{\itemsep}{0mm}%
\usecounter{enumi}}
 \def\newblock{\hskip .11em plus .33em minus .07em}
 \sloppy\clubpenalty4000\widowpenalty4000
 \sfcode`\.=1000\relax}

\renewcommand{\thefootnote}{\fnsymbol{footnote}}

\def \invisible{\mbox{$\rule{0mm}{1mm}$}}
\def \mathbox(#1){\invisible\ifmmode{{#1}}\else{\mbox{${#1}$}}\fi}
\def \mbf(#1){\mbox{\boldmath{$#1$}}}

\def\III{{\rm I\thinspace I\thinspace I}}

\def\etal{{\it et al.}}

\def\ie{{\it i.e.}}

\def\piii{p_{\rm III}}

\newcommand{\bra}{\langle}
\newcommand{\half}{\mbox{$\frac{1}{2}$}}
\newcommand{\ket}{\rangle}

\def\VIII{\mathbox(V_{\rm III})}

\def\FRAC#1#2{\leavevmode\kern-.em
\raise.5ex\hbox{\the\scriptfont0 #1}\kern-.em
/\kern-.15em\lower.25ex\hbox{\the\scriptfont0 #2}}
\newif\ifrefphysrev

\def\refNP{\refphysrevfalse
           \typeout{** Reference: Nucl Phys format}}
\refphysrevtrue

\def \vol(#1,#2,#3){\ifrefphysrev{{\bf {#1}}, 
{#3} (19{#2})}\else{{{\bf {#1}}(19{#2}){#3}}}\fi}
\def \NP(#1,#2,#3){Nucl.\ Phys.\          \vol(#1,#2,#3)}
\def \PL(#1,#2,#3){Phys.\ Lett.\          \vol(#1,#2,#3)}
\def \PRL(#1,#2,#3){Phys.\ Rev.\ Lett.\   \vol(#1,#2,#3)}
\def \PRp(#1,#2,#3){Phys.\ Rep.\          \vol(#1,#2,#3)}
\def \PR(#1,#2,#3){Phys.\ Rev.\           \vol(#1,#2,#3)}
\def \PTP(#1,#2,#3){Prog.\ Theor.\ Phys.\ \vol(#1,#2,#3)}
\def \ibid(#1,#2,#3){{\it ibid.}\         \vol(#1,#2,#3)}
\def \dataA(#1,#2,#3,#4,#5,#6){\put(#4,#2){\framebox(#5,#3){}}\put(#4,#1){\makebox(#5,-6)[c]{\bf ****}}
     \put(#4,#2){\makebox(#5,0)[t]{\rule{0pt}{10pt}{\small \sf #6}}}}
\def \dataB(#1,#2,#3,#4,#5,#6){\put(#4,#2){\framebox(#5,#3){}}\put(#4,#1){\makebox(#5,-6)[c]{\bf ***}}
     \put(#4,#2){\makebox(#5,0)[t]{\rule{0pt}{10pt}{\small \sf #6}}}}
\def \dataC(#1,#2,#3,#4,#5,#6){\put(#4,#2){\framebox(#5,#3){}}\put(#4,#1){\makebox(#5,-6)[c]{\bf **}}
     \put(#4,#2){\makebox(#5,0)[t]{\rule{0pt}{10pt}{\small \sf #6}}}}
\def \dataD(#1,#2,#3,#4,#5,#6){\put(#4,#2){\framebox(#5,#3){}}\put(#4,#1){\makebox(#5,-6)[c]{\bf *}}
     \put(#4,#2){\makebox(#5,0)[t]{\rule{0pt}{10pt}{\small \sf #6}}}}
\def \dataqC(#1,#2,#3,#4,#5,#6){\put(#4,#1){\makebox(#5,-6)[c]{\bf **}}
     \put(#4,#1){\makebox(#5,0)[t]{\rule{0pt}{10pt}{\small \sf #6}}}}
\def \dataqD(#1,#2,#3,#4,#5,#6){\put(#4,#1){\makebox(#5,-6)[c]{\bf *}}
     \put(#4,#1){\makebox(#5,0)[t]{\rule{0pt}{10pt}{\small \sf #6}}}}
\def \data(#1,#2,#3,#4,#5,#6){\put(#4,#2){\framebox(#5,#3){}}\put(#4,#1){\dashbox{5}(#5,0){}}
     \put(#4,#2){\makebox(#5,0)[t]{\rule{0pt}{10pt}{\small \sf #6}}}}
\def \dataq(#1,#2,#3,#4,#5,#6){\put(#4,#1){\dashbox{5}(#5,0){}}
     \put(#4,#1){\makebox(#5,0)[t]{\rule{0pt}{10pt}{\small \sf #6}}}}
\def \calc(#1,#2,#3){\put(#2,#1){\line(1,0){30}}
     \put(#2,#1){\makebox(30,0)[b]{\vspace*{0.5ex}{\small \sf #3}}}}
\def \calcb(#1,#2,#3){\put(#2,#1){\line(1,0){30}}
     \put(#2,#1){\makebox(30,0)[t]{\rule{0pt}{10pt}{\small \sf #3}}}}
\def \calco(#1,#2,#3){\put(#2,#1){\line(1,0){25}\raisebox{-0.2em}{{\sf{o}}}}
     \put(#2,#1){\makebox(30,0)[b]{\vspace*{0.5ex}{\small \sf #3}}}}
\def \calcbo(#1,#2,#3){\put(#2,#1){\line(1,0){25}\raisebox{-.2em}{\sf{o}}}
     \put(#2,#1){\makebox(30,0)[t]{\rule{0pt}{10pt}{\small \sf #3}}}}
\def \calcx(#1,#2,#3){\put(#2,#1){\line(1,0){25}\raisebox{-0.2em}{{\sf{x}}}}
     \put(#2,#1){\makebox(30,0)[b]{\vspace*{0.5ex}{\small \sf #3}}}}
\def \calcbx(#1,#2,#3){\put(#2,#1){\line(1,0){25}\raisebox{-.2em}{\sf{x}}}
     \put(#2,#1){\makebox(30,0)[t]{\rule{0pt}{10pt}{\small \sf #3}}}}

\def \chiru(#1,#2,#3){%
     \put(#2,#1){\makebox(30,0){\sf #3}}
      \put(#2,#1){\rule{1.2ex}{0pt}\circle{30}} }

\makeatletter
\def\scriptsize{\@setsize\scriptsize{14.5pt}\xipt\@xipt
\abovedisplayskip 11\p@ plus3\p@ minus6\p@
\belowdisplayskip \abovedisplayskip
\abovedisplayshortskip  \z@ plus3\p@
\belowdisplayshortskip  6.5\p@ plus3.5\p@ minus3\p@
\def\@listi{\leftmargin\leftmargini
\parsep 4.5\p@ plus2\p@ minus\p@ \itemsep \parsep
\topsep 9\p@ plus3\p@ minus5\p@}}
\def \@magscale#1{ scaled \magstep #1}
\font\frtnsfb = cmssbx10 \@magscale2 
\def \half(#1){\mathbox(\frac{#1}{2})}
\def \ninej(#1,#2,#3,#4,#5,#6,#7,#8,#9){\mathbox(\left\{\matrix 
     {#1&#2&#3\cr#4&#5&#6\cr#7&#8&#9\cr}\right\})}
\newif\ifnoncomplete

\noncompletefalse

\def\@cite#1#2{\unskip\nobreak\relax
    {[#1]}} 

\def\citenum#1{{\def\@cite##1##2{##1}\cite{#1}}}
\def\citea#1{\@cite{#1}{}}
 
\newcount\@tempcntc
\def\@citex[#1]#2{\if@filesw\immediate\write\@auxout{%
\string\citation{#2}}\fi
  \@tempcnta\z@\@tempcntb\m@ne\def\@citea{}\@cite{\@for\@citeb:=#2\do
    {\@ifundefined
       {b@\@citeb}{\@citeo\@tempcntb\m@ne\@citea\def\@citea{,}%
{\bf ?}\@warning
       {Citation `\@citeb' on page \thepage \space undefined}}%
{\setbox\z@\hbox{\global\@tempcntc0\csname b@\@citeb\endcsname\relax}%
     \ifnum\@tempcntc=\z@ \@citeo\@tempcntb\m@ne
       \@citea\def\@citea{,}\hbox{\csname b@\@citeb\endcsname}%
     \else
      \advance\@tempcntb\@ne
      \ifnum\@tempcntb=\@tempcntc
      \else\advance\@tempcntb\m@ne\@citeo
      \@tempcnta\@tempcntc\@tempcntb\@tempcntc\fi\fi}}\@citeo}{#1}}
\def\@citeo{\ifnum\@tempcnta>\@tempcntb\else\@citea\def\@citea{,}%
  \ifnum\@tempcnta=\@tempcntb\the\@tempcnta\else
   {\advance\@tempcnta\@ne\ifnum\@tempcnta=\@tempcntb %
\else \def\@citea{--}\fi
    \advance\@tempcnta\m@ne\the\@tempcnta\@citea\the\@tempcntb}\fi\fi}

\def\affiliation#1{\gdef\@affiliation{#1}}
 
\def\and{\cr \makebox[0in]{\rule[-1cm]{0mm}{1cm}and } \cr}

\def\maketitle{\par
 \begingroup
 \def\thefootnote{\fnsymbol{footnote}}
 \def\@makefnmark{\hbox
 to 0pt{$^{\@thefnmark}$\hss}}
 \if@twocolumn
 \twocolumn[\@maketitle]
 \else \newpage
 \global\@topnum\z@ \@maketitle \fi\thispagestyle{plain}\@thanks
 \endgroup
 \setcounter{footnote}{0}
 \let\maketitle\relax
 \let\@maketitle\relax
 \gdef\@thanks{}\gdef\@author{}\gdef\@title{}
 \gdef\@affiliation{} \let\affiliation\relax	%
 \let\thanks\relax}

\def\@maketitle{\newpage
 \null
 \vskip 0em plus 2em minus 0em     
 \ifx\@date\@empty\else
   \begin{flushright}
    {\ifnoncomplete(\today)
     \else{{\normalsize \@date}\\}\fi}      
   \end{flushright}
   \vskip 3em plus 2em minus 2em   
 \fi
 \begin{center}
  {\Large \@title \par}     
  \vskip 3em plus 1em minus 1.5em  
  {
   \lineskip .5em plus 0em minus .3em   
   \begin{tabular}[t]{c}\@author\\
   \end{tabular}\par}
  \vskip 0.5em plus 1em minus 1.5em  
  { \sl \@affiliation \par}
\end{center}
 \par
 \vskip 6em plus 2em minus 4em}     
 
\def\abstract{\if@twocolumn
\section*{Abstract}
\else \normalsize
\fi}
 
\def\endabstract{\if@twocolumn\fi\par\clearpage}

 \makeatother
\refNP
   \input epsf           
\begin{document}
{\frtnsfb
\noindent
Symmetric and Antisymmetric Spin-Orbit Forces \smallskip \\ 
\noindent
in YN Interaction 
 by a Quark Model
\footnotetext{
Talk at the 1st SUT-KEK Seminar on 6 Apr 1998. E-mail:
sachiko@thaxp1.tanashi.kek.jp}
}
\\

\noindent
Sachiko Takeuchi
\\
{\sl Dept of Public Health and Env Sci,\\
Tokyo Medical and Dental Univ,\\
 Yushima, Bunkyo, Tokyo 113-8519, Japan}
\\
\medskip

The symmetric and antisymmetric spin-orbit forces (SLS and ALS) in the 
YN interaction 
are investigated 
for relative $P$-wave systems
by a valence quark model 
with the instanton-induced interaction (\III).
The size of the adiabatic potential at the zero range is shown
for each of the YN channels.
The size of ALS is comparable to SLS.
The channel dependence of ALS, which
is determined by the flavor SU(3) symmetry 
when the one-gluon exchange (OGE) and/or the meson 
exchange interaction
are used, deviates after introducing \III.
In most of the two-baryon channels,
including the two-nucleon channel,
the spin-orbit force of the YN interaction 
is strong.
A few exceptional channels, however, 
are found where  \III\ and OGE are canceled to each other, 
and the spin-orbit force becomes small \cite{Ta98}.
\medskip

Recent experiments on the systems with strangeness are making great progress.
Especially the gamma spectroscopy has identified several gamma transitions, 
which give us valuable information 
on the spin part of the $\Lambda$N interaction
\cite{exp}.
From the observed levels of $\Lambda$-hypernuclei, 
it is believed that the spin-orbit force 
between $\Lambda$ and nucleon is very small 
comparing to that between two nucleons.
However, it is nontrivial to remove the nuclear effects.
Also, only the combined effect of LS and ALS can be measured in the hypernuclei.
Information on the noncentral parts of the YN interaction 
has not given directly from experiments yet.
The theoretical investigation of hypernuclei
has been performed mainly by using the
empirical YN interactions \cite{theoM}.
Here we employ a valence quark model 
to investigate the properties of the spin-orbit force in the strange systems.
The quark model with \III\
is found to have an appropriate size and the channel dependence 
for the spin-orbit force and therefore
will enable us to see
the feature from a more fundamental viewpoint.

 A valence quark model usually 
contains three terms in the hamiltonian: 
the kinetic term, the confinement term, and the
OGE term 
\cite{Is92,theoNNTokyo,theoYNTokyo,theoYNKyoto}.
It is considered that OGE stands for the
perturbative gluon effects and that the confinement force represents 
the long-range nonperturbative gluon effects.
We argue that a valence quark model 
should include \III\ as a
short-range nonperturbative gluon effect 
in addition to the other gluon effects
\cite{Ta98,OT91,Ta94}.
The model hamiltonian for quarks can be written as follows:
\begin{eqnarray}
H_{\rm quark} &=& K+(1-\piii)V_{\rm OGE} + \piii \VIII\ + V_{\rm conf}\, ,
\label{eq1}
\end{eqnarray}
where $V_{\rm OGE}$ and \VIII\
are the Galilei invariant terms of the \III\ and OGE potentials.
The parameter $\piii$ 
represents the relative strength of the
spin-spin part of \III\ to OGE.

The QCD instantons were originally introduced in  relation to 
the $U_A$(1) problem.
It produces couplings of instantons to the surrounding 
light-quark zero modes
\cite{tH76}.
This leads a flavor-singlet interaction among quarks,
which is \III\ in the present model.
This interaction is considered to 
be the origin of the
observed large mass difference of $\eta'$-$\eta$ mesons.

It is well known 
that the color magnetic interaction in OGE 
is responsible to produce many of the hadron properties.
By adjusting the strength of OGE,
it can reproduce
the hyperfine splittings (e.g., ground state N-$\Delta$ mass difference) 
as well as the short-range repulsion of
the two-nucleon systems in the relative $S$-wave 
\cite{Is92,theoNNTokyo}.
It, however, is also known that
the strength of OGE, $\alpha_s$, determined in this
empirical way is much larger than 1, which makes it hard 
to treat it as the perturbative effect.
The spin-spin part of 
\III\ produces the nucleon-$\Delta$ mass difference
and the short-range repulsion between
the two nucleons as the color magnetic interaction \cite{OT91,tH76}.  
Thus, one can reduces the
empirical strength of OGE,  $\alpha_s$, by introducing \III.
Moreover, LS of \III\ contributes the spin-orbit force in an interesting way
\cite{Ta98,Ta94}.

The valence quark model including only OGE 
as an origin of the hyperfine splittings
has a spin-orbit problem.
The LS part of OGE is strong; it is just strong enough 
to explain the observed large spin-orbit force between two nucleons 
\cite{theoNNTokyo}.
On the other hand, 
the experimental mass spectrum of the excited baryons indicates 
that such a strong spin-orbit force should not exist between quarks \cite{pdg}.
A valence quark model in which the spin-orbit parts 
of the quark 
 interaction are removed by hands 
can simulate the  observed mass spectrum 
\cite{Is92}. 
To explain both of the spin-orbit features at the same time 
is highly nontrivial.

In ref.\ \cite{Ta98,Ta94}, we demonstrated that 
introducing \III\ may solve the above
difficulty in the $P$-wave systems
due to  the channel-specific cancellation between
OGE and \III.
The mechanism of the cancellation 
is clearly seen at the flavor SU(3) limit.
Since here we consider the spin-orbit force on the $P$-wave systems,
the quark pairs which are orbital-antisymmetric and spin-symmetric,
\ie,
only the pairs symmetric (or antisymmetric) simultaneously 
in the flavor and
in the color spaces, are relevant.
It is found that the contribution from the color-symmetric
quark pairs dominates in most of  the YN LS interaction,
while only color-antisymmetric pairs exist in a baryon.
Because \III\ behaves like scalar-particle exchange
 and
OGE is vector-particle exchange,
the sign of their spin-orbit parts is opposite to each other.
Thus, where both of OGE and \III\ contribute,
namely for the color- and flavor-antisymmetric pairs,
LS cancellation occurs.
Therefore, it was expected 
that introducing \III\ 
would explain the strong LS in the two-nucleon systems
and weak LS in the excited baryons,
which was  confirmed numerically.

The negative-parity baryon mass spectrum by the present model
shows that there is only weak spin-orbit force between the quarks
due to the above cancellation.
In the present choice of the parameters,
the LS splittings becomes from 0.14 to 0.37 times smaller
than that from the model only with OGE (fig.\ 1) \cite{Ta98}.

$\rule{3cm}{0mm}$
\begin{minipage}{10cm}
\epsfbox{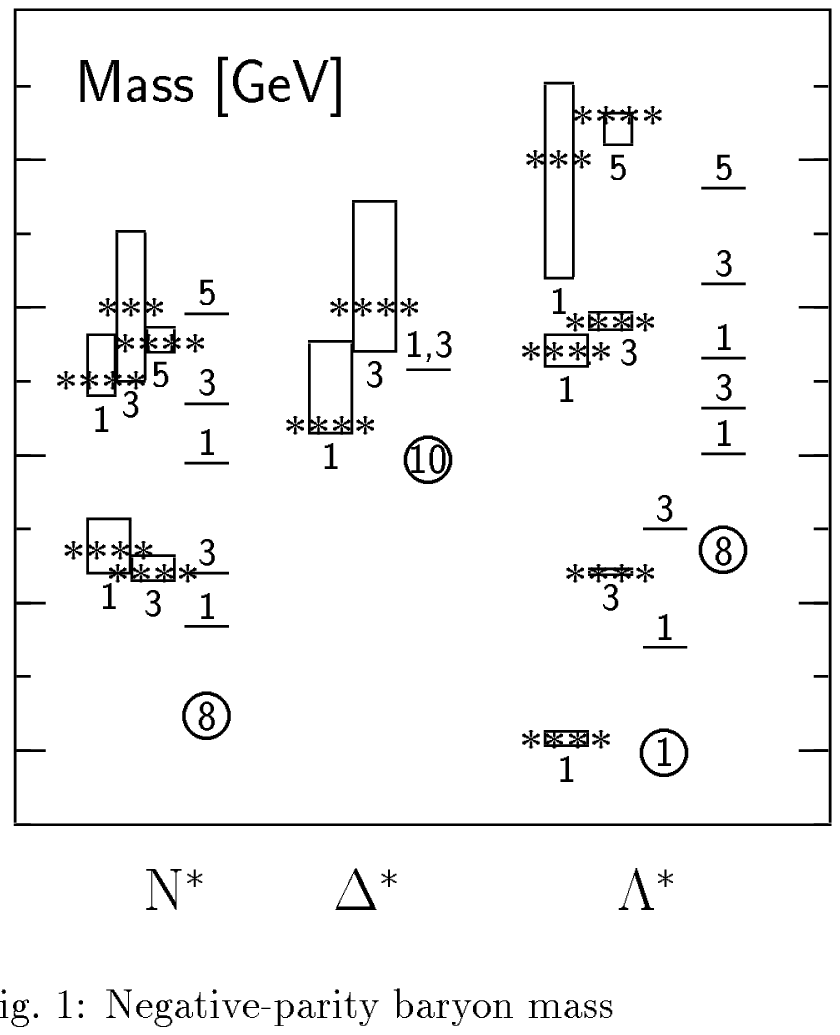}
\end{minipage}

The LS part of the quark interaction produces both of the 
LS and ALS terms in the YN interaction.
To investigate their feature, we calculate
an adiabatic potential for the two baryons at relative
distance, $R=0$.
It can be obtained by subtracting the single particle value
from the matrix element taken
by the $(0p)(0s)^{5}$ harmonic oscillator wave function:
\vfill

\noindent
\begin{minipage}{5cm}
\setlength{\parindent}{0.6cm}
\begin{eqnarray*}
\lefteqn{U(\makebox{$R$=0})=}&&\\
&& \bra \Psi | V | \Psi \ket
 - \sum_i^2 \bra \phi_i | V | \phi_i 
\ket .
\nonumber
\end{eqnarray*}
This value is considered to express
the size of the short-range interaction between the baryons.

The  numerical results of $U_{\rm SLS[ALS]}$($R$=0) at $\xi_s$ = 1 and 0.6 
with and without \III\ are listed in table 1.
The size of ALS is comparable to SLS in general.
Both kinds of the spin-orbit forces depends strongly 
on the channels.

\end{minipage}
\hfill
\begin{minipage}{11cm}
$\rule{0cm}{3mm}$
$\rule{10cm}{0.1mm}$\\
\noindent
Table 1: Spin-orbit forces 
in the two-baryon systems\\
The values are in MeV with the relative strength to NN SLS.
\\

\vspace*{-1mm}
\setlength{\tabcolsep}{0.5mm}
\noindent
\begin{tabular}{lccc|rr|rr|rr|rr}
\hline
\hline
	& &  & & \multicolumn{4}{c|}{$m_u/m_s=1$} & \multicolumn{4}{c}{$m_u/m_s=0.6$} \\
\cline{5-12}
	& 2$T$ & $S$ & $S'$  & \multicolumn{2}{c|}{$\piii=0$} & \multicolumn{2}{c|}{$\piii=0.4$} & 
\multicolumn{2}{c|}{$\piii=0$} & \multicolumn{2}{c}{$\piii=0.4$}\\
\hline
SLS\\
\hline
NN-NN                 & 2 & 1 & 1 &  $-$94 &    1.00 & $-$61 &    1.00 & $-$94 &    1.00 & $-$61 &    1.00 \\
N$\Lambda$-N$\Lambda$ & 1 & 1 & 1 &  $-$73 &    0.78 & $-$51 &    0.83 & $-$54 &    0.57 & $-$37 &    0.61 \\
N$\Sigma$-N$\Sigma$   & 1 & 1 & 1 &     22 & $-$0.24 &  $-$3 &    0.05 &    22 & $-$0.23 &  $-$4 &    0.06 \\
N$\Lambda$-N$\Sigma$  & 1 & 1 & 1 &  $-$32 &    0.34 & $-$14 &    0.22 & $-$28 &    0.30 & $-$13 &    0.21 \\
N$\Sigma$-N$\Sigma$   & 3 & 1 & 1 &  $-$94 &    1.00 & $-$61 &    1.00 & $-$96 &    1.02 & $-$61 &    0.99 \\
\hline
ALS\\
\hline
N$\Lambda$-N$\Lambda$ & 1 & 1 & 0 &  $-$37 &    0.39 & $-$18 &    0.30 & $-$37 &    0.40 & $-$23 &    0.38 \\
N$\Sigma$-N$\Sigma$   & 1 & 1 & 0 &     87 & $-$0.93 &    44 & $-$0.71 &    79 & $-$0.85 &    43 & $-$0.69 \\
N$\Lambda$-N$\Sigma$  & 1 & 1 & 0 &  $-$37 &    0.39 & $-$18 &    0.30 & $-$30 &    0.32 & $-$19 &    0.31 \\
N$\Lambda$-N$\Sigma$  & 1 & 0 & 1 &     87 & $-$0.93 &    44 & $-$0.71 &    79 & $-$0.84 &    42 & $-$0.69 \\
N$\Sigma$-N$\Sigma$   & 3 & 1 & 0 &      0 &    0.00 &     0 &    0.00 &  $-$1 &    0.01 &     3 & $-$0.05 \\
\hline
\hline
\end{tabular}
\end{minipage}

The symmetric and antisymmetric spin-orbit force between two baryons
 remains strong
in most of the channels after introducing \III\ as was found 
in the two-nucleon system.
There, however, are a few exceptional channels where 
color-antisymmetric quark pairs play an important role
and the OGE-\III\ LS 
cancellation also
gives notable effects:
the symmetric spin-orbit force of the N$\Sigma(I=1/2)$ and 
of the N$\Lambda$-N$\Sigma$
channels becomes small after introducing \III.

Introducing \III\
changes the channel dependence of the spin-orbit force.
It was reported that if the interaction between baryons 
holds the flavor SU(3),
which corresponds to the $\piii$ = 0 and $\xi_s$ = 1 case here, 
the channel dependence is determined only by the SU(3) symmetry 
\cite{theoALS}.
The $U_{\rm SLS[ALS]}(R)$ from qSLS at $\piii$=0 
were calculated and found to hold 
the above relation between the channels
except for the factor from the norm kernel \cite{theoALS}.
Since \III\ affects color- and flavor-antisymmetric quark pairs selectively, 
this relation  deviates when \III\ is switched on.
We found that introducing \III\ changes actually 
the relative strength of the spin-orbit force between the baryons 
considerably.

The adiabatic potential at $R>0$ looks like a 
gaussian with the range of about 1 fm \cite{theoYNTokyo,theoALS}.
Since the potential we are considering here
is SLS or ALS between relative $P$-wave, 
the potential at $R>0$ 
will be more important.
Moreover, when we treat the quarks dynamically by, {\it e.g.},
 a quark cluster model,
the potential we should consider between baryons
is not
the adiabatic one but the RGM potential, which is highly
nonlocal \cite{theoNNTokyo,theoYNTokyo,theoYNKyoto} 
(fig.\ 2)\footnotetext{The values for ALS in fig.\ 2 have been revised.}.
We argue, however, as far as a relative strength of SLS or ALS 
 to the NN SLS is concerned, the conclusion here
holds even when
one performs more 
sophisticated calculations.
\begin{figure}
\epsfbox{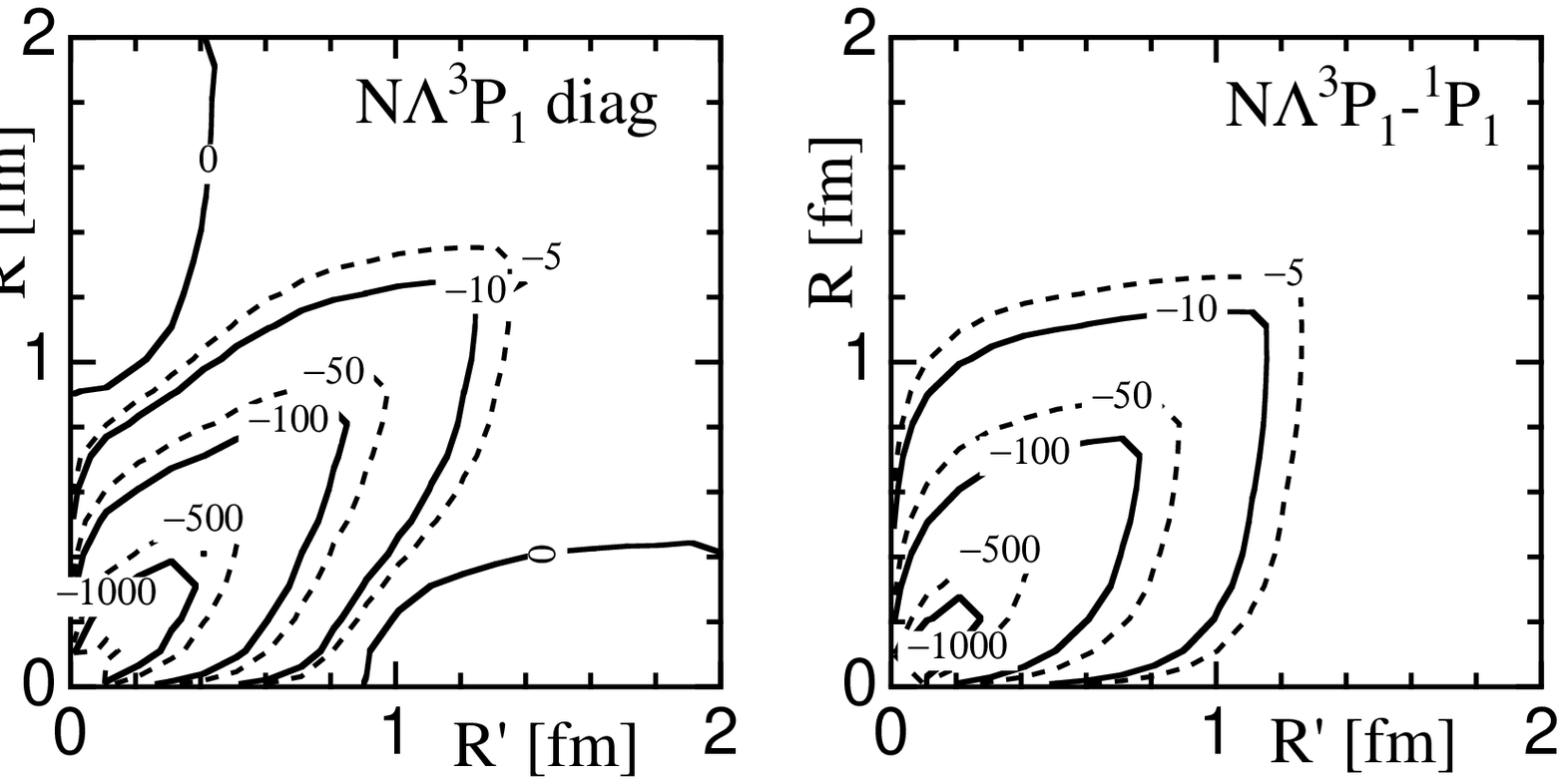}
\begin{center}
Fig.\ 2: LS and ALS parts of the RGM kernel
\end{center}
\end{figure}

The ALS term is also in the meson-exchange interaction \cite{theoALS};
it was found that the tensor couplings of the vector-meson exchange 
can produce ALS at the flavor SU(3) limit. 
The meson-induced ALS seems much smaller than
that of quarks, though the size of the meson coupling is not well known
\cite{theoALS}.

Experimentally, information on the YN spin-orbit force has been given only 
through the level splittings in $\Lambda$ hypernuclei.
The level splittings, however, gives only the combined strength
of SLS and ALS; the strong SLS and ALS originated from 
the quark interaction may cancels each other.
It was reported that the other effect such as the YN tensor interaction
may reduce the splitting \cite{theoM}.
More investigations both from the theoretical and experimental sides
are necessary 
to understand the spin properties of the systems with strangeness.

\smallskip

The author would like to thank K.\ Yazaki and M.\ Oka 
for valuable discussions.
This work was supported in part by the Grant-in-Aid for scientific research 
Priority Areas (Strangeness Nuclear Physics) of the Ministry of Education, 
Science, Sports and Culture of Japan.

\begin{thebib}{99}
\bibitem{Ta98}
Part of this work is reported in S.\ Takeuchi, hep-ph/9807240.

\bibitem{exp}
S.\ Ajimura, Proceedings of 1st SUT-KEK seminar (Apr, 1998);
K.\ Tanida,  Proceedings of 1st SUT-KEK seminar (Apr, 1998).

\bibitem{theoM}
C.\ B.\ Dover and H.\ Feshbach, Ann.\ Phys.\ (N.Y.) \vol(198,90,321);
\ibid(217,92,51);
H.\ Hiyama, \etal, Proceedings of 1st SUT-KEK seminar (Apr, 1998).

\bibitem{Is92} 
N.\ Isgur, In.\ J.\ of Mod.\ Phys., \vol(1,92,465) 
and references therein;
G.\ Karl, In.\ J.\ of Mod.\ Phys., \vol(1,92,491).

\bibitem{theoNNTokyo}  
O.\ Morimatsu, K.\ Yazaki, and M.\ Oka, \NP(A424,84,412); 
S.\ Takeuchi, K.\ Shimizu, and K.\ Yazaki, \NP(A504,89,777);
K.\ Shimizu, Rep.\ Prog.\ Phys.\ \vol(52,89,1) 
and references therein.

\bibitem{theoYNTokyo}  
O.\ Morimatsu, S.\ Ohta, K.\ Shimizu, and K.\ Yazaki, \NP(A420,84,573); 
Y.\ Koike, \NP(A454,86,509);
M.\ Oka, Prog.\ Theor.\ Phys.\ Suppl. \vol(120,95,95).

\bibitem{theoYNKyoto}
Y.\ Fujiwara, C.\ Nakamoto, and Y.\ Suzuki, \PR(C54,96,2180);
\PRL(76,96,2242) and references therein.

\bibitem{OT91}
M.\ Oka and S.\ Takeuchi, \NP(A524,91,649); 
\PRL(63,89,1780);
S.\ Takeuchi and M.\ Oka, \PRL(66,91,1271).

\bibitem{Ta94}
S.\ Takeuchi,  \PRL(73,94,2173);
%
S.\ Takeuchi,  \PR(D53,96,6619).

\bibitem{tH76}
G.\ 't Hooft, \PR(D14,76,3432);
M.A.\ Shifman, A.I.\ Vainshtein and V.I.\ Zakharov, \NP(B163,80,46);
E.V.\ Shuryak, \PRp(C115,84,151);
%
N.I.\ Kochelev, Sov.\ J.\ Nucl.\ Phys.\ \vol(41,85,291);
E.V.\ Shuryak and J.L.\ Rosner,  \PL(B218,89,72).

\bibitem{pdg}
C.\ Caso, \etal, The European Physical Journal \vol(C3,98,1)

\bibitem{theoALS}
M.\ Oka, \NP(A629,98,379c); 
M.\ Oka and Y.\ Tani, in preparation.

\end{thebib}

\end{document}